\documentclass[reprint, superscriptaddress, amsmath,amssymb,prb,twocolumn]{revtex4}
\usepackage[utf8]{inputenc}
\usepackage{epsfig}
\usepackage{graphicx}
\usepackage{adjustbox}
\usepackage{indentfirst}
\usepackage{multirow}
\usepackage{latexsym}
\usepackage{color, soul}
\usepackage{ragged2e}
\usepackage{upgreek}
\usepackage{amsmath}
\usepackage{amssymb}
\usepackage{mathtools}

\usepackage{hyperref}
\usepackage{dcolumn}
\usepackage{graphicx, xcolor}
\usepackage{sansmath}
\usepackage{float}

\newcounter{bla}

 \usepackage[normalem]{ulem}
\def\bk{\ensuremath{{\mathbf{k}}}}
\def\bq{\ensuremath{{\mathbf{q}}}}
\def\br{\ensuremath{{\mathbf{r}}}}
\def\sgw{\mbox{\textsc{SternheimerGW}}}
\def\GW{GW}
\def\GWnot{G$_0$W$_0$}
\def\Ex{\ensuremath{{E_{\rm x}}}}
\def\Ec{\ensuremath{{E_{\rm c}}}}
\def\me{\ensuremath{{m_{\rm e}}}}
\def\mh{\ensuremath{{m_{\rm h}}}}
\def\mr{\ensuremath{{m_{\rm r}}}}
\def\Sr{\ensuremath{\Sigma}}
\def\gap{QP gap}

\makeatletter
\DeclareMathOperator{\myhelper@Re}{Re} 
\renewcommand{\Re}{\myhelper@Re}
\DeclareMathOperator{\myhelper@Im}{Im} 
\renewcommand{\Im}{\myhelper@Im}
\makeatother

\begin{document}

\title{GW band structure of monolayer MoS$_2$ using the SternheimerGW method and effect of dielectric environment}

\author{Nourdine Zibouche}
\thanks{To whom correspondence should be addressed:~n.zibouche@bath.ac.uk}
\affiliation{Department of Chemistry, University of Bath, Bath BA2 7AY, United Kingdom\\}
\author{ Martin Schlipf}
\affiliation{VASP Software GmbH, Sensengasse 8/12, A-1090 Vienna, Austria\\}
\author{Feliciano Giustino}
\thanks{To whom correspondence should be addressed:~fgiustino@oden.utexas.edu}
\affiliation{Oden Institute for Computational Engineering and Sciences, The University of Texas at Austin, Austin, 
Texas 78712, USA}
\affiliation{Department of Physics, The University of Texas at Austin, Austin, Texas 78712, USA}

\begin{abstract}
Monolayers of transition-metal dichalcogenides (TMDs) hold great promise as future nanoelectronic and optoelectronic devices. An essential feature for achieving high device performance is the use of suitable supporting substrates, which can affect the electronic and optical properties of these two-dimensional (2D) materials.  Here, we perform many-body \GW{} calculations using the SternheimerGW method to investigate the quasiparticle band structure of monolayer MoS$_2$ subject to an effective dielectric screening model, which is meant to approximately describe substrate polarization in real device applications. We show that, within this model, the dielectric screening has a sizable effect on the quasiparticle band gap, for example the gap renormalization is as large as 250 meV for MoS$_2$ with model screening corresponding to SiO$_2$. Within the \GWnot{} approximation, we also find that the inclusion of the effective screening induces a direct band gap, in contrast to the unscreened monolayer. We also find that the dielectric screening induces an enhancement of the carrier effective masses by as much as 27\% for holes, shifts plasmon satellites, and redistributes quasiparticle weight. Our results highlight the importance of the dielectric environment in the design of 2D TMD-based devices.

\end{abstract}

\maketitle

\section{Introduction}
Semiconducting compounds of layered transition-metal dichalcogenides (TMDs) in their two-dimensional (2D) forms have exceptional properties. They undergo an indirect-to-direct band gap transition in the monolayer limit,\cite{Mak2010,Kuc2011}  and they exhibit a strong spin-orbit coupling,\cite{Zibouche2014a} and tightly bound excitons\cite{He2014,Steinhoff2018,Waldecker2019,Deilmann2019,Gjerding2020} and trions,\cite{Mak2013,Florian2018,Goswami2019} which give rise to interesting spin-valley physics.\cite{Cao2012,Refaely-Abramson2018, Yong2019,Zhang2019} They also offer the possibility of designing a variety of van der Waals heterostructures.\cite{Yu2013,Utama2019}
In the past decade there have been significant advances in the synthesis and fabrication\cite{Chhowalla2013} of TMDs, opening up many opportunities in applications for nanoelectronics and optoelectronics, including photodetectors,\cite{Lopez-Sanchez2013} lasers,\cite{Salehzadeh2015} light emitting diodes,\cite{Withers2015} memory devices,\cite{Roy2013} sensors,\cite{Zhang2015} and field-effect transistors.\cite{Radisavljevic2011,Yu2013a}
Two-dimensional TMDs exhibit strong Coulomb interactions associated with the weak dielectric screening in two dimensions.\cite{Cudazzo2011,Berkelbach2013} Consequently, the polarization of the supporting substrate modifies electron-electron and electron-hole interactions, thus renormalizing the quasiparticle gap and reducing the exciton binding energies. For example, the measured electronic band gap on a SiO$_2$ substrate is 2.10~eV,\cite{Zhou2016,Kerelsky2017,Goryca2019,Klein2019} whereas values of 1.9~eV\cite{Park2018} and 2.40~eV\cite{Huang2015} have been reported on gold and graphite substrates, respectively. The exciton binding energy spans a wide range, between 0.2~eV and 0.9~eV,\cite{Cheiw2012,Klein2019,Cao2015,Park2018,Qiu2013,Klots2014,Goryca2019} depending on the substrate.
Several experimental and theoretical studies reported substrate-dependent electronic and optical properties of these atomically thin TMDs, such as variations in the carrier mobilities and transport properties,\cite{Radisavljevic2013,Bao2013,Liu2013,Yu2016,Huo2018} exciton binding energies and lifetimes,\cite{Korn2011,Cheiw2012,Shi2013a,Lin2014,Cao2015,Palummo2015} luminescence efficiency,\cite{Sercombe2013,Buscema2014,Scheuschner2014,Yu2016a} and band gap renormalization.\cite{Zhang2014,Bruix2016,Rigosi2016,Zhou2016,Kerelsky2017,Klein2019}
This sensitivity to the substrate calls for an investigation of the role of environmental screening in the electronic properties of 2D materials. 

Previous studies in this field focused on the effect of the substrate on the band gap and the binding energies.\cite{Hueser2013, Lin2014,Ryou2016,Park2018,Steinhoff2018,Waldecker2019}
Since calculations with explicit substrates to capture the screening of a semi-infinite bulk insulator are currently beyond reach, all previous work relied on simple models of substrate screening. Here we also model the substrate screening using an effective dielectric continuum, and we expand on previous work by investigating the effect of dielectric screening on the quasiparticle bands, carrier effective masses, spectral density and plasmon satellites.
To this aim, we perform state-of-the-art many-body \GW{} calculations for the archetypal TMD monolayer MoS$_2$. Substrate polarization is accounted for within a simple model whereby we screen the Coulomb potential entering the calculation of the polarizability within the random phase approximation. To make the analysis directly relevant to experiments, we choose the dielectric constants corresponding to  hexagonal boron nitride (h-BN) and SiO$_2$, which are commonly used with TMDs. We show that the renormalization of quasiparticle energies can be significant; for example the band gap of monolayer MoS$_2$ decreases by as much as 250~meV when considering a SiO$_2$ substrate, and the hole effective mass increases by 27\%. Furthermore, we find that the model dielectric environment changes the nature of the gap from indirect to direct, 
and shifts plasmon satellites.

This paper is organized as follows. In Sec.~\ref{sec.methods} we briefly review the SternheimerGW method used in this work, we discuss computational details, and we provide numerical convergence tests. In Sec.~\ref{sec.results} we report our results on the quasiparticle band structure of monolayer MoS$_2$ in the presence of an effective dielectric screening, we analyze the renormalization of the band gap and effective masses, and we discuss the influence of the dielectric environment on the spectral function and plasmon satellites. In Sec.~\ref{sec.conclusions} we summarize our findings and offer our conclusions.
\section{Methods}\label{sec.methods}
\subsection{The Sternheimer\GW{} method}

\begin{figure*}
\begin{center}
 \includegraphics[width=\textwidth]{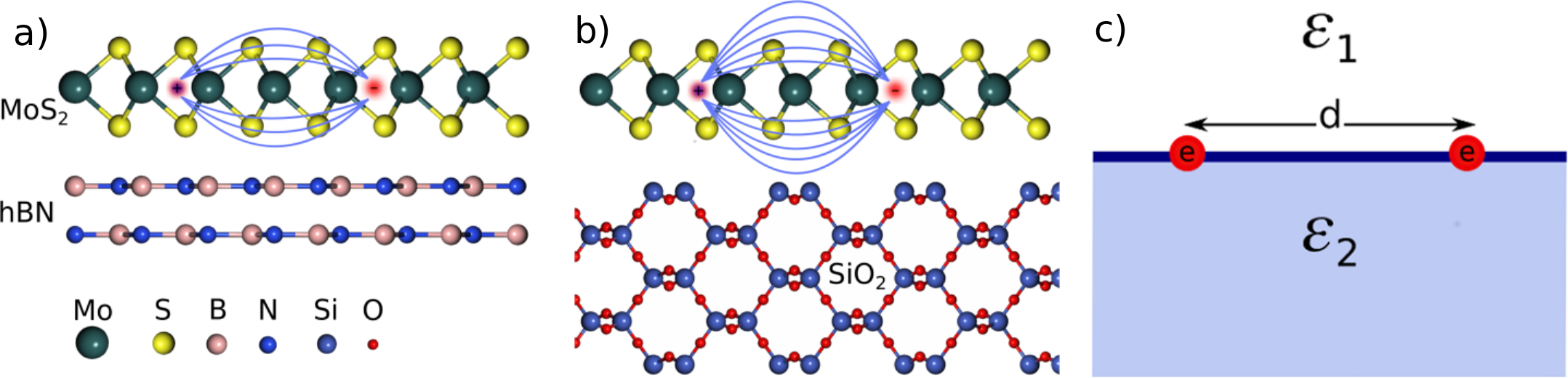}
   \caption{(a) and (b), Schematic of a MoS$_2$ monolayer on an h-BN and an SiO$_2$ substrates. In the present work the substrate is modeled using an effective dielectric environment, as shown in (c). The Coulomb interaction between two point charges with charge $e$ sitting at the distance $d$ right at the interface is $e^2/4\pi\varepsilon_0 \varepsilon_{\rm eff}$, where $\varepsilon_0$ is the permittivity of vacuum and $\varepsilon_{\rm eff} = (\varepsilon_1+\varepsilon_2)/2$. The derivation of this result can be found in Sec.~4.4 of Ref.~\onlinecite{jackson_classical_1999}, among others.}
 \label{fig:Schematic}
\end{center}
\end{figure*}

The \GW{} method\cite{Hedin1965,Hybertsen1986,Onida2002,Golze2019} has emerged as the most successful \textit{ab initio} approach for calculating many-body quasiparticle band structures in semiconductors.
The method is based on the calculation of the electron self-energy, which includes exchange and correlation effects via the dynamically screened Coulomb interaction. The screened Coulomb interaction $W$ is most often calculated within the random-phase approximation (RPA), starting from Kohn-Sham wavefunctions obtained within density functional theory (DFT) \cite{dft_kohn_sham}.

Standard implementations of the \GW{} methods obtain the electron Green's function and the RPA polarizability by using an expansion over unoccupied Kohn-Sham states.\cite{Kresse1996,Deslippe2012,Gonze2020}
Although very successful, in this approach the convergence with respect to unoccupied states is challenging, which results in a heavy computational load.
To circumvent this bottleneck, several groups have been pursuing direct calculations of $G$ and/or $W$ using the Sternheimer equation or variants of this method.\cite{Baroni1987,Baroni2001,Umari2009,Umari2010,Govoni2015}
In this work we employ the SternheimerGW method that we developed,~\cite{Schlipf2020} in which both the screened Coulomb interaction and the electron Green's function are evaluated using solely the occupied Kohn-Sham states. Below we briefly review this methodology. More details and the derivation of key equations can be found in Refs.~\onlinecite{Schlipf2020,Lambert2013,Giustino2010}.

The Green’s function $G(\br, \br'; \omega)$ and the screened Coulomb interaction $W(\br, \br'; \omega)$ are expressed in terms of the space coordinate $\br'$, while $\br$ and $\omega$ are treated as parametric space and frequency variables.
The Green’s function is calculated by solving the inhomogeneous linear system of equations for all occupied states
\begin{equation} \label{eq:green}
       (\hat H - \hbar \omega)~G_{0[\br, \omega]}(\br') = -\delta_{\br -\br'}.
\end{equation}
Here, $\hat H$ corresponds to the single-particle Kohn-Sham Hamiltonian and $\delta$ is the Dirac delta function.

The screened Coulomb interaction $W_0(\br, \br'; \omega)$ within the RPA\cite{Bohm1951,Pines1952,Bohm1953} can be obtained with the procedure outlined below.
When the system is subject to a perturbation $\Delta V_{[\br, \pm\omega]}(\br')$, the corresponding change in the charge density is given by
\begin{equation} 
        \Delta n_{[\br, \omega]}(\br') = 2\sum_{\nu}\psi^*_\nu(\br') [\Delta\psi_{{\nu}[\br, + \omega]}(\br') + \Delta\psi_{\nu[\br, -\omega]}(\br')],
 \end{equation}
where $\Delta\psi_{\nu[\br, \pm\omega]}(\br')$ are the frequency-dependent variations of the occupied single-particle wavefunctions. These variations are obtained by solving the following Sternheimer equation
\begin{equation}
 (\hat H - \epsilon_\nu \pm\hbar\omega)\Delta\psi_{\nu[\br, \pm\omega]}(\br') = -(1 - \hat P_\nu)\Delta V_{[\br, \pm\omega]}(\br')\psi_{\nu}(\br').
\end{equation}
The operator $\hat P_{\rm v} = \sum_\nu^{\rm occ.} \lvert \psi_{\nu}\rangle\langle\psi_{\nu}\rvert$
projects onto the occupied manifold, and $\epsilon_\nu$ are the corresponding Kohn-Sham energy eigenvalues.
There are two methods for choosing the perturbation $\Delta V_{[\br, \pm\omega]}(\br')$ that yield $W_0(\br, \br'; \omega)$. In the direct (non-self-consistent) approach, the perturbation is set to the bare Coulomb potential $\Delta V_{[\br, \pm\omega]}(\br')~=~v(\br, \br')$.
From the variation in the charge density, the RPA dielectric function is evaluated as
\begin{equation}
     \varepsilon_{[\br, \omega]}(\br') = \delta_{\br - \br'} - \Delta n_{[\br, \omega]}(\br').
\end{equation}
The screened Coulomb interaction $W_0(\br, \br'; \omega)$ is then calculated by inverting $\varepsilon$ via
\begin{equation}
    W_{0[\br, \omega]}(\br') = \int d\br''v(\br, \br'')  \varepsilon^{-1}(\br'',\br'; \omega).
\end{equation}
In the self-consistent method, the perturbation is set to the screened Coulomb interaction $\Delta V_{[\br, \pm\omega]}(\br') = W_0(\br, \br'; \omega$). This scheme initializes the perturbation $\Delta V_{[\br, \pm\omega]}(\br')$ to the bare Coulomb interaction $v(\br, \br')$. Then, the induced variation in the charge density $\Delta n_{[\br, \omega]}(\br')$ generates a Hartree potential that screens the bare Coulomb interaction through
\begin{equation}
    \Delta V_{[\br, \omega]}(\br') = \int d\br''\Delta n_{[\br, \omega]}(\br'')v(\br'', \br') .
\end{equation}
The updated screened Coulomb interaction $W_0(\br, \br'; \omega)$, given by
\begin{equation}
    W_{0[\br, \omega]}(\br') =  v(\br, \br') + \Delta V_{[\br, \omega]}(\br'),
\end{equation}
is subsequently used to evaluate the next density response. This process is iterated until convergence is reached.

The self-energy, $\Sigma$, is obtained as the product of the Green's function $G_0$ and the screened Coulomb interaction $W_0$ 
\begin{equation} 
   \Sigma(\br, \br'; \omega') = \frac{i}{2\pi} \int\limits_{-\infty}^{+\infty} G_0(\br, \br'; \omega + \omega') W_0(\br, \br'; \omega')e^{-i\delta\omega'}d\omega',
\end{equation}
and the quasiparticle energies can thus be determined as 
\begin{equation}
    \epsilon_{n\bk}^{QP} =  \epsilon_{n\bk} + Z_{n\bk}\langle\psi_{n\bk}\lvert\Sigma(\epsilon_{n\bk})-V^{\rm xc}_{n\bk} \lvert\psi_{n\bk}\rangle,
\end{equation}
where $\epsilon_{n\bk}$, $\psi_{n\bk}$, and $V^{\rm xc}_{n\bk}$ are, respectively, the Kohn-Sham DFT eigenvalues, wavefunctions, and the expectation value of the exchange-correlation potential of the $n^{\rm th}$ band. $Z_{n\bk}~=~[1~-~\langle\psi_{n\bk}\lvert\partial{\Sigma(\epsilon)}/\partial{\epsilon}\lvert_{\epsilon=\epsilon_{n\bk}} \lvert\psi_{n\bk}\rangle]^{-1}$ is the quasiparticle renormalization factor that defines the quasiparticle weight carried by the excitation.
 The SternheimerGW method provides the possibility of calculating the complete energy- and momentum-resolved spectral function $A(\omega, \bk)$, a physical observable that can be extracted from angle-resolved photoemission spectroscopy (ARPES) measurements. $A(\omega, \bk)$ is calculated as

\begin{multline}
 A(\omega, \bk)=\frac{1}{\pi}\times\\
 \sum_{n}\frac{|\Im\Sigma_n(\omega,\bk)|}{[\omega-\epsilon_{n\bk}-\Delta\Re\Sigma_n(\omega,\bk)]^2+[\Im\Sigma_n(\omega,\bk)]^2},
\end{multline} 

%

in which $\Re\Sigma$ and $\Im\Sigma$ indicate the real and imaginary parts of the \GWnot{} self-energy, and $\Delta \Re\Sigma_n(\omega,\bk) = \Re\Sigma_n(\omega,\bk)-V^{\rm xc}_{n\bk}$.

\subsection{Computational details}

\begin{figure*}
\begin{center}
 \includegraphics[width=\textwidth]{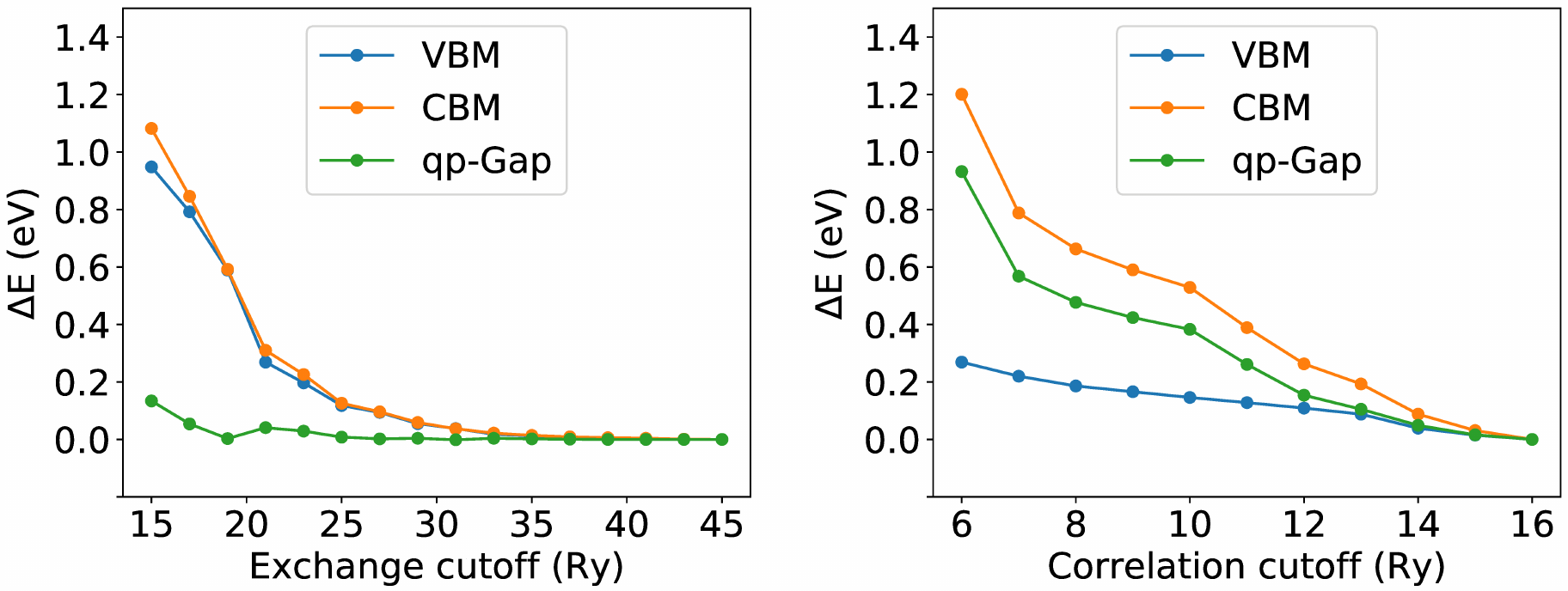}
  \caption{Difference ($\Delta E$) of the quasiparticle band gap (\gap{}), valence band maximum (VBM) and conduction band minimum (CBM) from the corresponding converged values, as a function of (a) exchange ($\Ex$) and (b) correlation ($\Ec$) self-energy cutoff. The values are obtained at the high symmetry K~point. The differences between the last two values of the gap in (a) and (b) are 2~meV and 16~meV, respectively.}
\label{fig:qp-Gap-vbm-cbm}
\end{center}
\end{figure*}

Ground-state calculations are carried out using density functional theory as implemented in the Quantum ESPRESSO  package.\cite{qespresso2017,Giannozzi2009} The Kohn-Sham wavefunctions and energies are calculated using the Perdew-Burke-Ernzerhof (PBE) functional.\cite{Perdew1996} A plane-wave basis is used with energy and charge-density cutoffs of 50 Ry and 200 Ry, respectively. We approximate the core-valence interactions via norm-conserving pseudopotentials, explicitly including the 4$s$ and 4$p$ semicore electrons of Mo. The Brillouin zone (BZ) integration is sampled using a $15\times15\times1$ Monkhorst-Pack $\bk$-point grid.\cite{Monkhorst1976} The atomic positions are relaxed at the experimental lattice constant $a$ = 3.16~\AA. To avoid spurious interactions between periodically repeated slabs, the size of the computational cell, including monolayer and vacuum, is set to 20~\AA{} in the out-of-plane direction, unless otherwise stated.

We perform \GWnot{} calculations starting from the PBE wavefunctions and energy eigenvalues.
The dielectric matrix $\varepsilon$ is computed within the random phase approximation using either the Godby-Needs plasmon-pole approximation (PPA)\cite{Godby1989} with an imaginary pole energy of 16~eV, or using full frequency integration (FF), as implemented in the \sgw~code.\cite{Schlipf2020,Lambert2013,Giustino2010} The FF integration is performed along the imaginary axis using 65 discrete frequencies in the interval of 0~eV to 240~eV.
We obtain the FF self-energy on the real axis using an analytic continuation following the adaptive Antoulas-Anderson method.\cite{Nakatsukasa2018}

To avoid spurious Coulomb interactions between electrons belonging to periodic images of the monolayer, we truncate the Coulomb interaction $v$, in the calculation of both the dielectric function, $\varepsilon$, and the screened Coulomb interaction, $W = \varepsilon^{-1} v$. In particular, we employ a 2D truncation scheme in reciprocal space, using the expression from Refs.~\onlinecite{Rozzi2006} and \onlinecite{Ismail-Beigi2006}: $v_{\rm 2D}(\bk)=4\pi[1-\exp(-\sqrt{k^2_x+k^2_y}L_z)\cos(k_z L_z)]/|{\bf k}|^2$. Here $L_z$ is the cutoff distance in the out-of-plane direction. At the DFT level, we truncate the bare Coulomb potential using the scheme of Ref.~\onlinecite{Sohier2017}, which speeds up the convergence of the \GW{} calculations with respect to the Brillouin zone grid. We note that this truncation is important: without truncation the \GW{} band gap would be underestimated by about 0.26~eV.

In order to take into account the effect of substrate polarization, we renormalize the screened Coulomb interaction by the effective background dielectric constant ($\varepsilon_{\rm eff}$) through
\begin{equation}
 \varepsilon_{\rm eff} = \left ( 1 + \varepsilon_{\rm s} \right )/2,
 \label{eq:diel_func}
\end{equation}

where $\varepsilon_{\rm s}$ refers to the relative dielectric constant of the substrate.\cite{Hwang2007,Polini2008,Hwang2008}

This effective dielectric constant is obtained by evaluating the Coulomb interaction between two point charges at the (mathematically sharp) interface between vacuum and a semi-infinite dielectric continuum, as shown in Fig.~\ref{fig:Schematic}.~\cite{jackson_classical_1999}
Using this approach we model two substrate materials, SiO$_2$ ($\varepsilon_{\rm s}=3.9$)\cite{Robertson2004} and a monolayer or a few layers of h-BN ($\varepsilon_{\rm s}=2.6$).\cite{Hyder1976,Kim2012}
Figure.~\ref{fig:Schematic} shows a qualitative schematic of the systems that we model, however we emphasize that our calculations contain a single layer of MoS$_2$, without substrate atoms.

\subsection{Numerical convergence tests}

\begin{table}
\centering
\caption{Dependence of the quasiparticle band bap (\gap{}), valence band maximum (VBM) and conduction band minimum (CBM) at the high symmetry {K}~point on the number of $\bq$-points used to sample the BZ. The exchange ($\Ex$) and correlation ($\Ec$) self-energy cutoffs are set to 45~Ry and 15~Ry, respectively.}
\label{my-kpoints}
\begin{ruledtabular}
\begin{tabular}{ccddd}
$\bq$ mesh & Irreducible $\bq$ points & \multicolumn{1}{c}{VBM} & \multicolumn{1}{c}{CBM} & \multicolumn{1}{c}{\gap{}}  \\
\hline
$09\times09\times1$ &  12  & -5.864 & -2.822 & 3.04 \\
$12\times12\times1$ &  19  & -5.729 & -2.949 & 2.78 \\
$15\times15\times1$ &  27  & -5.726 & -3.006 & 2.72 \\
$18\times18\times1$ &  37  & -5.753 & -3.038 & 2.72 \\
$21\times21\times1$ &  48  & -5.785 & -3.054 & 2.73 \\
    \end{tabular}
\end{ruledtabular}
\end{table}

\begin{figure*}
\begin{center}
 \includegraphics[width=\textwidth]{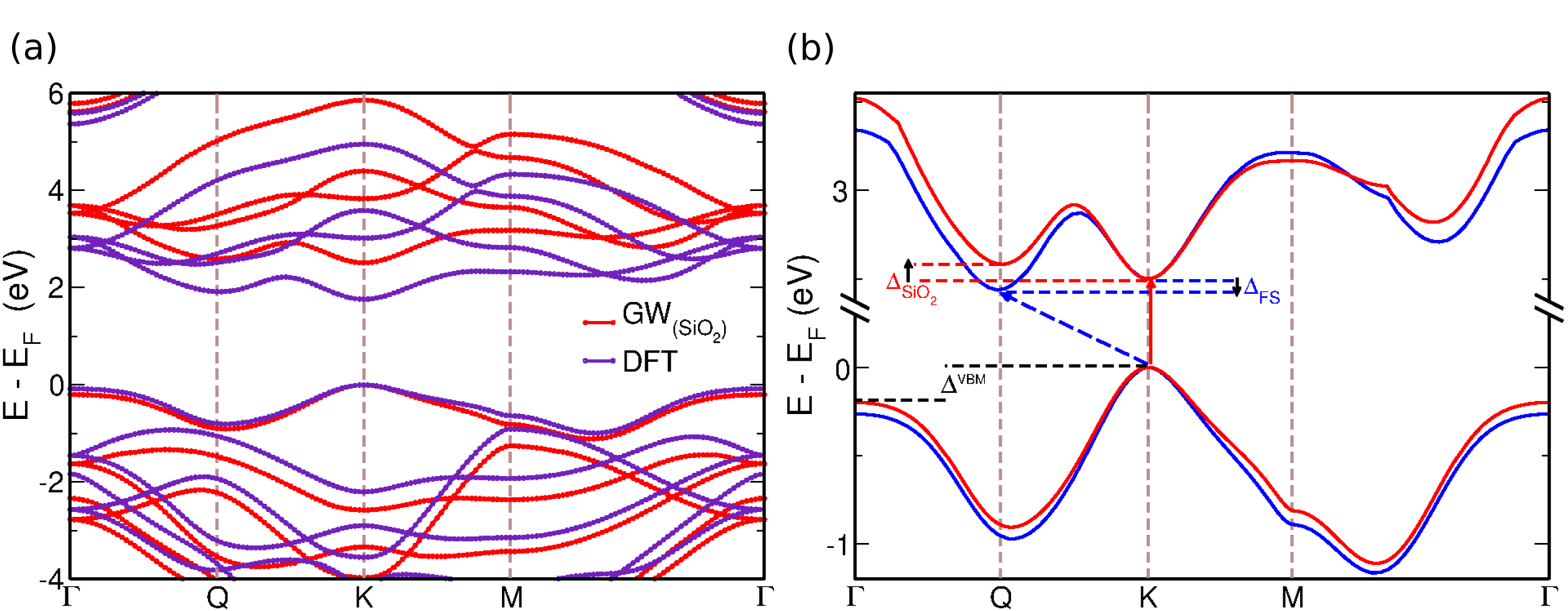}
 \caption{(a) \GWnot{} (red) and DFT (indigo) band structures of monolayer MoS$_2$. The origin of the energy axis is set to the VBM at the {K}~point. The \GWnot{} band structures are calculate within the PPA, using the dielectric screening model corresponding to a SiO$_2$ substrate. (b) Highest valence band and lowest conduction band calculated within \GWnot{}/PPA, highlighting the change in the band gap character from indirect to direct when moving from the unscreened case [``freestanding'' (FS) blue] to screening by an SiO$_2$ substrate (red). The CBM at {K} have been aligned for comparison.}
 \label{fig:gw_dft_bands}
\end{center}
\end{figure*}

For accurate results \GW{} calculations require the convergence of several numerical parameters. In this section, we discuss the dependence of the band gap and the energy of the band extrema with respect to the energy cutoff for exchange and correlation, as well as the sampling of the Brillouin zone using the PPA.

First, we focus on the convergence of the quasiparticle band gap (\gap{}) and the valence band maximum (VBM) and conduction band minimum (CBM) at the $K$ point, with respect to the exchange ($\Ex$) and the correlation ($\Ec$) energy cutoffs. The relative changes compared to the converged values are shown in Fig.~\ref{fig:qp-Gap-vbm-cbm}. To study the convergence of $\Ex$ we set a correlation cutoff $\Ec = 15~\text{Ry}$ (1 Ry = 13.605 eV); conversely, to study the convergence with respect to $\Ec$, the exchange cutoff is set to $\Ex = 45~\text{Ry}$. In both convergence tests, the BZ is sampled using a $15\times15\times1$ $\mathbf q$-point mesh (27 irreducible points) for the dielectric matrix and the screened Coulomb interaction. Figure.~\ref{fig:qp-Gap-vbm-cbm}{a} shows that VBM and CBM are well converged for $\Ex$ above 35~Ry, increasing by only 12 meV when we increase the cutoff all the way to 45 Ry. Since both band extrema converge from the top at a similar rate, the \gap{} converges much faster, and is accurate to within 2~meV already for $\Ex = 25~\text{Ry}$. Figure.~\ref{fig:qp-Gap-vbm-cbm}{b} shows that the convergence with respect to $\Ec$ is somewhat slower, but the changes in the VBM, CBM, and \gap{} from $\Ec =15$ Ry to 16 Ry are of 15 meV, 31 meV, and 16 meV respectively. For $\Ec = 16~\text{Ry}$, the \gap{} is found to be 2.70~eV, which is remarkably (and probably coincidentally) the same value as reported in experiments on suspended layers.\cite{Lin2014}

Next, we focus on the number of grid points used for sampling the BZ to evaluate the dielectric matrix and the screened Coulomb interaction within the PPA. Table~\ref{my-kpoints} reports the convergence of the \gap{}, the VBM, the CBM at the high-symmetry {K} point at fixed energy cutoffs $\Ex = 45~\text{Ry}$ and $\Ec = 15~\text{Ry}$. A $\mathbf q$-point grid of $15\times15\times1$ is necessary to converge the self-energy with 50~meV accuracy. The resulting \gap{} at the K point is in very good agreement with previous \GW{} calculations, yielding 2.60--2.80~eV.\cite{Cheiw2012,Shi2013,Hueser2013,Jin2015a} As in the present work, these previous calculations employ the experimental lattice parameter. Differences between reported band gaps arise from differences in the \GW{} calculations,  specifically the Coulomb truncation and the vacuum size. Despite such differences, our calculations also indicate that the \GWnot{} band gap of a pristine MoS$_2$ monolayer is indirect.
We do not include spin-orbit coupling in our calculations, because the resulting energy splittings at the K point amount to 3~meV (CBM) and 147~meV (VBM) at the DFT level, which is below the numerical precision of our \GW{} calculations.

\section{Results and Discussion}\label{sec.results}

\subsection{Quasiparticle band gap and band structures}

In this section we discuss our results for the quasiparticle band gap and band structure of monolayer MoS$_2$, as obtained by considering a layer in vacuum, the effective screening from a SiO$_2$ substrate, and the effective screening resulting from h-BN. The following results correspond to exchange and correlation cutoffs $\Ex = 45~\text{Ry}$ and $\Ec = 15~\text{Ry}$, and a $15\times15\times1$ $\bq$-point grid.
In Fig.~\ref{fig:gw_dft_bands}{a}, we compare the DFT and the \GWnot{}/PPA band structure of a MoS$_2$ monolayer in the presence of the model dielectric screening corresponding to SiO$_2$. The \GWnot{} correction is not uniform throughout the Brillouin zone, so that not only the band gap but also the effective masses are modified (see Sec.~\ref{sec:eff_mass}). 
In Table~\ref{tab:gaps}, we compare the calculated \gap{}, VBM, and CBM at the high-symmetry K point of the unscreened [``freestanding'' (FS)] MoS$_2$ monolayer, with a monolayer in the presence of screening from an effective substrate with the dielectric constants of h-BN or SiO$_2$.
When using full frequency integration, the band extrema shift to lower energies, and the \gap{} is reduced compared to that in the PPA model. This reduction ranges from 40~meV for the unscreened monolayer to 80 meV for the screened monolayer.

We find that the model substrate screening renormalizes the absolute quasiparticle energies of the VBM and CBM. As a consequence, the quasiparticle band gap is also reduced as compared to the unscreened monolayer. In particular, we find a reduction of the band gap by 180(140) meV when using FF(PPA) frequency integration for h-BN, and of 250(210) meV for SiO$_2$.
This reduction is consistent with the notion that the Coulomb energy required for adding/removing an electron in monolayer MoS$_2$ should be reduced by the dielectric screening of the substrate.

\begin{figure}
\begin{center}
 \includegraphics[width=\columnwidth]{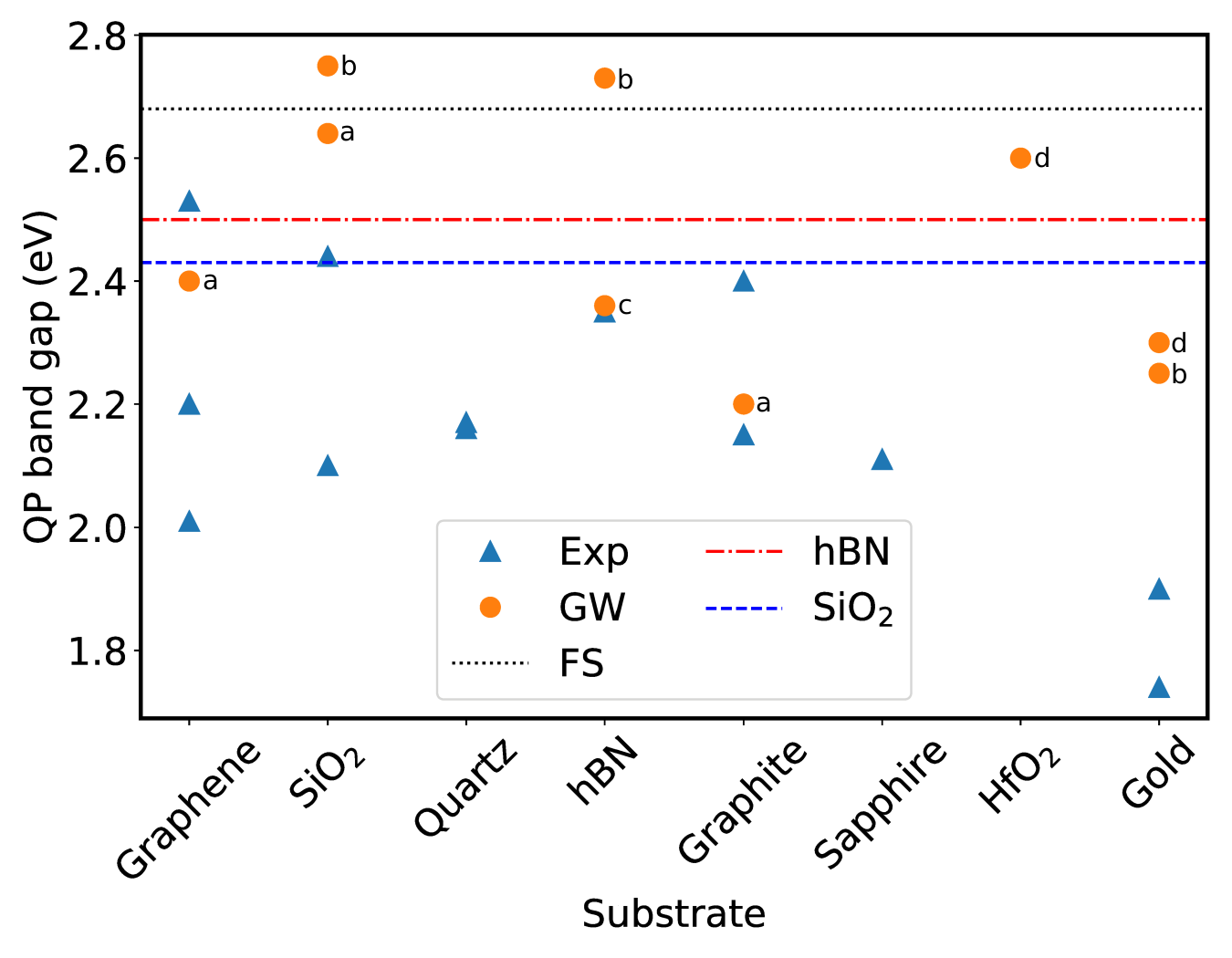}
 \caption{Quasiparticle band gap of a MoS$_2$ monolayer on different substrates reported in the literature. The experimental results shown by blue triangles were obtained with scanning tunneling microscopy/spectroscopy,\cite{Zhang2014,Lu2015,Huang2015,Chiu2015,Bruix2016,Liu2016,Rigosi2016,Hill2016,Shi2016,Zhou2016,Kerelsky2017,Murray2019,Klein2019} absorbance,\cite{Klein2019} and angle-resolved (inverse) photoemission spectroscopy (ARPES/ARIPES).\cite{Park2018} The \GW{} band gaps are shown by the orange disks.\cite{Ryou2016,Drueppel2017,Naik2018,Utama2019} The horizontal lines represent our calculated quasiparticle gaps using FF integration for the unscreened monolayer [``freestanding'' (FS)], and for h-BN and SiO$_2$ screening. The GW calculations of (a) Ref.~\onlinecite{Naik2018}, (b) Ref.~\onlinecite{Drueppel2017}, (c) Ref.~\onlinecite{Utama2019}, and (d) Ref.~\onlinecite{Ryou2016} reported in the plot were also been obtained by considering models to describe the effective environmental dielectric screening.}
\label{fig:Lit_gaps}
\end{center}
\end{figure}

In line with our finding, previous experimental and theoretical work indicates the sensitivity of the \gap{} to the dielectric screening environment, as shown in Fig.~\ref{fig:Lit_gaps}.
In the case of the SiO$_2$ substrate, scanning tunneling spectroscopy (STS) measurements obtain a \gap{} of 2.1~eV.\cite{Zhou2016,Kerelsky2017,Klein2019} However, optical absorption measurements on the same sample used for STS in Ref.~\onlinecite{Klein2019} yield a gap of 2.44~eV. This latter value agrees with our FF \gap{} (2.43~eV) for monolayer MoS$_2$ with model screening corresponding to SiO$_2$. Ref.~\onlinecite{Klein2019} argues that the tunneling gap is underestimated due to band-tail states near the conduction band minimum. Overall, the calculated band gaps from the literature, which we reproduce in  Fig.~\ref{fig:Lit_gaps}, are in qualitative agreement with experiments. However, the magnitude of the \gap{} renormalization is generally underestimated.
Particularly good agreement between theory and experiments is found for the MoS$_2$ monolayer on a h-BN substrate. The carefully converged \GW{} \gap{} (2.36~eV) from Ref.~\onlinecite{Utama2019} is very similar to the \gap{} measured by STS (2.35~eV) in Ref.~\onlinecite{Klein2019}. In our calculations, when we consider FF integration and $\varepsilon_{\rm s}=5.9$ corresponding to the dielectric constant of bulk h-BN, We obtain a \gap{} of 2.35~eV, which is in excellent agreement with the above theoretical and experimental values.

\begin{table}
\centering
\caption{Quasiparticle band bap (\gap{}, eV), valence band maximum (VBM) and conduction band minimum (CBM) at the high-symmetry {K}~point, for the unscreened [``free-standing, (FS)] MoS$_2$ monolayer, and for the same layer in the dielectric environment corresponding to an h-BN or a SiO$_2$ substrate.}
\label{tab:gaps}
\begin{ruledtabular}
\begin{tabular}{c dd dd dd}
  at {K} point & \multicolumn{2}{c}{VBM} & \multicolumn{2}{c}{CBM} & \multicolumn{2}{c}{\gap{}} \\
\cline{2-3} \cline{4-5} \cline{6-7}
  Substrate & \multicolumn{1}{c}{PPA} & \multicolumn{1}{c}{FF} & \multicolumn{1}{c}{PPA} & \multicolumn{1}{c}{FF} & \multicolumn{1}{c}{PPA} & \multicolumn{1}{c}{FF} \\
  \hline
  FS       & -5.726 & -5.905   & -3.006 & -3.233   & 2.72 & 2.68 \\
  h-BN      & -5.651 & -5.794   & -3.071 & -3.295   & 2.58 & 2.50 \\
  SiO$_2$  & -5.713 & -5.809   & -3.201 & -3.378   & 2.51 & 2.43 \\
  \end{tabular}
  \end{ruledtabular}
\end{table}

One interesting result of our calculations is that the screening of the substrate changes the character of the \gap{}. As mentioned above, \GWnot{} predicts an indirect \gap{} for the free-standing MoS$_2$ monolayer at the experimental lattice parameter (3.16~\AA). When we employ a model dielectric screening with the dielectric constants of SiO$_2$ or h-BN, we find a direct \gap{}. Figure.~\ref{fig:gw_dft_bands}{b} illustrates this change in between the free-standing monolayer and a monolayer in the presence of model dielectric screening corresponding to an SiO$_2$ substrate.
In the presence of substrate screening, the CBM at the midpoint, Q, of the high-symmetry $\Gamma$--{K} path (see Fig.~\ref{fig:gw_dft_bands}) raises above the CBM at the {K} point 
as compared to the unscreened case. Introducing the energy difference $\Delta$ =  CBM$_{\rm K}$ - CBM$_{Q}$, we find $\Delta_{\rm FS} = 98~\text{meV}$, $\Delta_{\rm hBN} = -57~\text{meV}$, and $\Delta_{\rm SiO_2} = -94~\text{meV}$ using the FF method. In the PPA calculations these differences are less pronounced: $\Delta_{\rm FS} = 65~\text{meV}$, $\Delta_{\rm hBN} = -15~\text{meV}$, and $\Delta_{\rm SiO_2} = -60~\text{meV}$. 
This result indicates that the screening-induced renormalization is more significant at the {K} point, and especially so when using FF integration. Unlike the CBM, the maximum of the valence band remains at the {K} point irrespective of substrate screening. The energy difference $\Delta^{\rm VBM}$ between the VBMs at the {K} and $\Gamma$ point are 0.23~eV, 0.19~eV and 0.17~eV for the unscreened, the h-BN-screened, and the SiO$_2$-screened monolayer, respectively. Again the PPA yields smaller differences, in the range of 20--30~meV. It should be noted that to predict a direct band gap at the \GWnot{} level, a full geometry relaxation (lattice parameters and atomic positions) is needed. Furthermore, self-consistent GW calculations also lead to direct band gap in the MoS$_2$ monolayer\cite{Cheiw2012,Shi2013} as observed in photoluminescence measurements.\cite{Mak2010,Lin2014}

Overall, the present results show that the dielectric environment alters qualitatively and quantitatively the \gap{} of monolayer MoS$_2$. It is natural to expect the same behavior for other monolayer TMDs. In addition to the effect of dielectric screening from a uniform semi-infinite substrate, which we consider here, it is expected that several other effects will contribute to renormalizing QP levels in these systems, for example the atomic-scale structure of the TMD/substrate interface, and the possible presence of interface dipoles, strain, moir\'es, and charge transfer. These effects should be considered by performing calculations using explicit substrates. The advantage of the simple model adopted here is that it includes long-range electrostatic effects that would not be captured by calculations using a substrate slab of finite thickness.

\subsection{Electron and hole effective masses} \label{sec:eff_mass}

The effective masses $m^* = \hbar^2 ( \partial^2E / \partial k^2 )^{-1}$
of electrons and holes at the {K} point are calculated along the high-symmetry {K--$\Gamma$} and {K--M} lines. We evaluate the second derivatives of the band curvatures numerically, using a step $\Delta k = 0.01$~\AA$^{-1}$ around the {K} point.
Since in the SternheimerGW method the Green's function and the screened Coulomb interaction are computed separately, we can directly determine quasiparticle energies $E_{\bf k}$ for arbitrary ${\bf k}$-points, without using interpolation techniques. Our calculated effective masses are shown in Table~\ref{tab:masses}. The electron and hole effective masses obtained within DFT are 0.43~{$m_0$} and 0.52~{$m_0$}, respectively, consistent with previously reported values \cite{Zibouche2014,Zibouche2014a} ($m_0$ indicates the free electron mass).  In the DFT calculations we do not include the substrate screening effect, so the reported DFT mass is independent of substrate screening. The \GWnot{} effective masses for the unscreened MoS$_2$ monolayer are in a good agreement with previous \GW{} data available in the literature, in the range of {0.35--0.40~$m_0$}\cite{Cheiw2012,Shi2013,Molina-Sanchez2015,Qiu2013} for electrons and {0.39--0.49~$m_0$}\cite{Cheiw2012,Shi2013,Molina-Sanchez2015} for holes.
In the presence of model substrate screening, the effective masses are heavier than for the unscreened monolayer (see Table~\ref{tab:masses}). This is consistent with Fig.~\ref{fig:gw_dft_bands}{b}, where we see that band curvatures at the {K} point are more pronounced when considering screening from SiO$_2$. We find that, for the model with the screening corresponding to a h-BN (SiO$_2$) substrate, the electron effective mass $\me$ is enhanced by 5\% (8\%) whereas the hole effective mass $\mh$ increases by 17\% (27\%) with respect to the unscreened layer. As for the quasiparticle shifts, the effective mass enhancement due to the screening is more pronounced for calculations performed with FF integration rather than the PPA.

\begin{table}
\centering
\caption{Calculated electron and hole effective masses of the free-standing (FS) and substrate-screened MoS$_2$ monolayer at the {K}~point.}
\label{tab:masses}
\begin{ruledtabular}
\begin{tabular}{c ddd ddd}
  at {K} point  & \multicolumn{3}{c}{$\me / m_0$} & \multicolumn{3}{c}{$\mh / m_0$}  \\
  \cline{2-4} \cline{5-7}
  Substrate & \multicolumn{1}{c}{PPA} & \multicolumn{1}{c}{FF} & \multicolumn{1}{c}{DFT} & \multicolumn{1}{c}{PPA} & \multicolumn{1}{c}{FF} & \multicolumn{1}{c}{DFT} \\
  \hline
   FS      & 0.39 & 0.39 & 0.43    & 0.42 & 0.41 & 0.52 \\
   h-BN     & 0.40 & 0.41 & 0.43    & 0.45 & 0.48 & 0.52 \\
   SiO$_2$ & 0.40 & 0.42 & 0.43    & 0.46 & 0.52 & 0.52 \\
  \end{tabular}
\end{ruledtabular}
\end{table}

Effective masses have been measured for a MoS$_2$ monolayer separated from a MoS$_2$ bulk compound by intercalating potassium using angle-resolved photoemission spectroscopy (ARPES).\cite{Eknapakul2014} The extracted effective masses at the {K} point are {$\me = (0.67 \pm 0.08)~m_0$} and {$\mh = (0.60 \pm 0.08)~m_0$}. These values are significantly higher than in our calculations and previous theoretical work. The difference could originate from the heavy doping of the conduction band with electrons by the potassium intercalation, which would induce metallic screening.\cite{Eknapakul2014,Miwa2015} This interpretation is consistent with the fact that the gap extracted from ARPES is , $1.86 \pm 0.02$~eV, is significantly smaller than other measured optical gaps and calculated quasiparticle gaps (see Fig.~\ref{fig:Lit_gaps}).
We also note that our calculations do not take into account the intercalant and electron-phonon interactions, which can both contribute to modifying the effective masses.

Additional ARPES measurements of the hole effective mass on different substrates have been reported. Ref.~\onlinecite{Jin2015} measured the hole effective mass for a suspended monolayer {($\mh = 0.43~m_0$)} and for a monolayer on SiO${_2}$ {($\mh = 0.48~m_0$)}. Their findings are very close to our calculations. Ref.~\onlinecite{Miwa2015,Dendzik2015} reported a hole effective mass of {$0.55 \pm 0.08~m_0$} for a MoS$_2$ monolayer grown on a gold substrate. Larger values of the effective masses, {$\mh = (0.81 \pm 0.05)~m_0$}\cite{Kim2016} and {$\mh = (0.66 \pm 0.04)~m_0$}\cite{Fregnaux2016}, have been reported for MoS$_2$ grown on SiO$_2$ by chemical vapor deposition. Also in this case, the high doping level is expected to contribute an effective mass enhancement compared to exfoliated monolayers.\cite{Fregnaux2016}

Our calculated reduced electron-hole {effective masses, $\mr = \me \mh / (\me + \mh)$}, for the unscreened monolayer and for h-BN and SiO${_2}$ screening, are 0.20~$m_0$, 0.22~$m_0$, and 0.23~$m_0$, respectively. These values should be compared with the measured exciton's reduced mass $\mr = 0.27~m_0$, as obtained from magneto-optical spectroscopy experiments.\cite{Goryca2019} The slight difference may be due to the fact that, in the experiment, the MoS$_2$ monolayer is encapsulated between slabs of h-BN, therefore the screening is enhanced compared to our calculations. 

\subsection{Self-energy and spectral function}

\begin{figure*}
\begin{center}
 \includegraphics[scale=0.37]{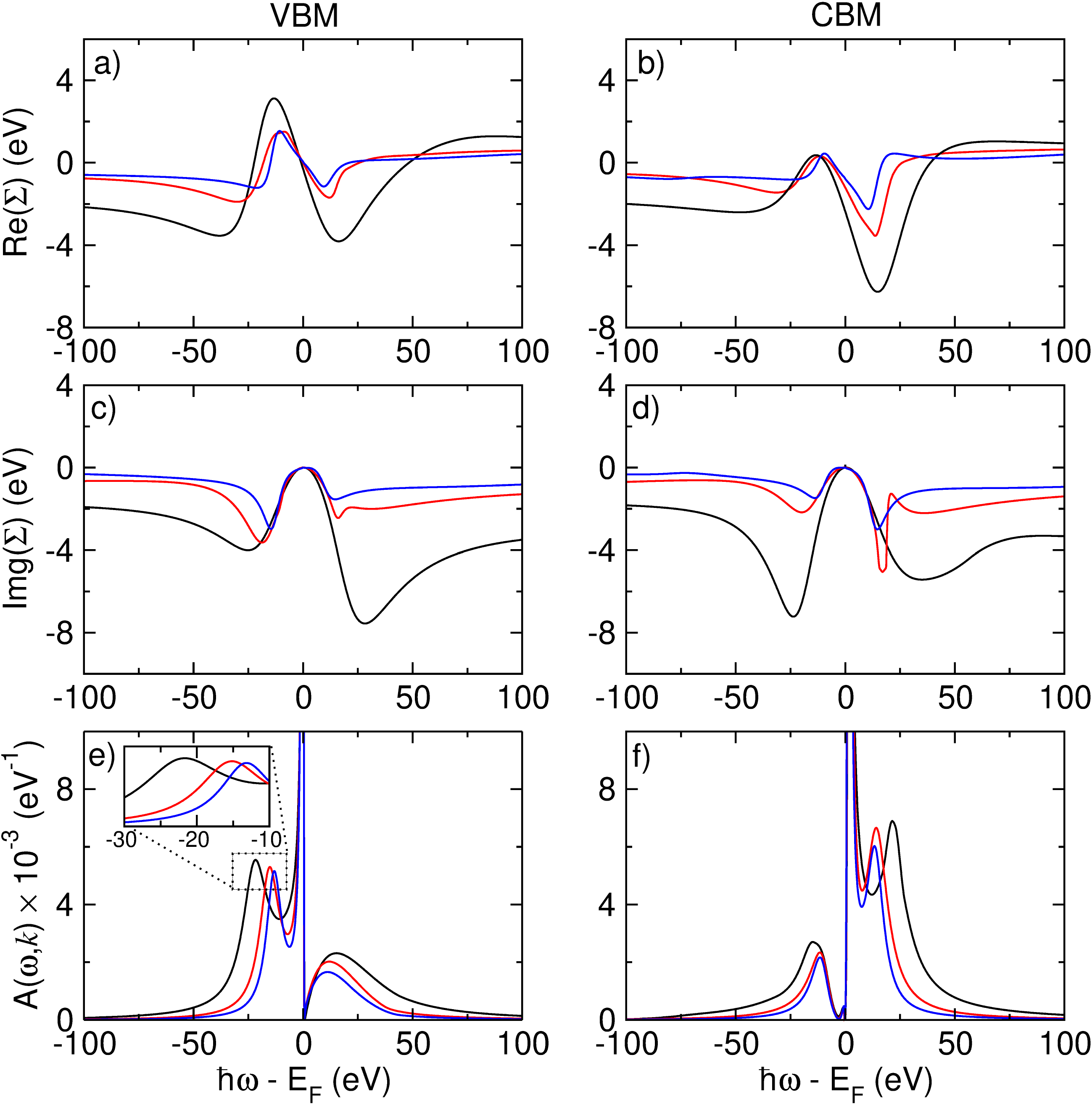}
  \caption{
   (a) and (b) Real part of the \GWnot{} self-energy ($\Sr$) of monolayer MoS$_2$ for the VBM and CBM states. (c) and (d) Corresponding imaginary part of the self energy. (e) and (f) Corresponding spectral functions $A(\omega,k$). All calculations are performed at the K point for the unscreened monolayer [``freestanding'' (FS), black], the case with h-BN screening (red), and the case with SiO$_2$ screening (blue).}
  \label{fig:real_sigma}
\end{center}
\end{figure*}

In this section, we discuss the effect of the dielectric screening on the electron self-energy, the spectral function, and the related incoherent plasmonic structure. For these calculations it is necessary to employ FF integration as opposed to the PPA. Figures.~\ref{fig:real_sigma}a-d show the frequency-dependent real and imaginary parts of the self-energy of the VBM and the CBM at the {K} point, for both the unscreened and screened MoS$_2$ monolayer. The real part determines the quasiparticle shift and renormalization, the imaginary part determines the quasiparticle broadening and lifetimes. We can see that both Re($\Sr$) and Im($\Sr$) have a pronounced structure in the range of 15--25~eV, which arises from plasmon excitations. In fact, the electron energy loss spectra of MoS$_2$ monolayer exhibit the characteristic of low-energy and high-energy plasmon resonances called $\pi$ and $\pi + \sigma$ at 7.6~eV and 15.6~eV, respectively, which arise from the collective excitation of the Mo $d$ and S $s,  p$ states.\cite{Liang1969,Johari2011} Here, the spectral function A($\omega, k$) in Figs.~\ref{fig:real_sigma}e,f clearly shows a plasmon satellite at around 22~eV, arising from the excitation of the high-energy $\pi + \sigma$ plasmons.\cite{Liang1969,Johari2011} On the other hand, the low-energy $\pi$ plasmons are not visible; these features possibly overlap with the broad main quasiparticle peaks. We emphasize that the energy and intensity of these plasmonic satellites are not captured correctly by \GWnot{}, which is known to overestimate the binding energy of satellites. For an accurate description of these features one would need to perform cumulant expansion calculations.\cite{Guzzo2012,Lischner2013,Guzzo2014, Kas2014, Caruso2015,Caruso2016} Earlier studies of plasmon satellites of TMDs within the cumulant expansion method can be found in Ref.~\onlinecite{Caruso2015}. 

When introducing substrate screening within the simplified model adopted in this work, these structures become less intense, and shift to lower binding energies.
This shift can be rationalized in terms of the Drude model, whereby the plasma frequency is given by $\omega_{\rm p} = \sqrt{ne^2 / \varepsilon_0m}$, where $n$, $e$ and $m$ are the electron density, charge and mass, respectively.\cite{Kittel2004} When substituting the permittivity of vacuum $\varepsilon_0$ with the effective dielectric constant of the substrate $\varepsilon_{\rm eff}$, the plasma frequency $\omega_{\rm p}^{\rm s}$ is reduced with respect to the unscreened monolayer, {$\omega_{\rm p}^{\rm s} = \omega_{\rm p}^{\rm FS} / \sqrt{\varepsilon_{\rm eff}}$}. The inset in Fig.~\ref{fig:real_sigma}e shows that our calculated shift in the plasma peaks is consistent with Drude's model. In fact, we find that the unscreened plasmon peak at 22~eV shifts to around 16~eV and 13~eV when we consider screening corresponding to h-BN and SiO$_2$ substrates, respectively.

From the real part of the self-energy we can evaluate the quasiparticle renormalization factors, $Z$. For the
VBM/CBM states at $K$ we find $Z=0.75$/0.77, 0.79/0.83, and 0.80/0.87 for the unscreened, and h-BN- and SiO$_2$-screened monolayers, respectively. These values indicate a weakly correlated electron system. The larger values associated with the larger screening are consistent with a lower transfer of quasiparticle weight to the plasmon satellites, and hence reduced correlations, as can be seen in the spectral function plots in Figs.~\ref{fig:real_sigma}e,f    .

\section{Conclusions}\label{sec.conclusions}
In summary, we investigated the dielectric screening effect of a substrate on the quasiparticle properties of monolayer MoS$_2$ using the first-principles SternheimerGW method and a simplified effective dielectric model to account for substrate polarization. We showed that the additional screening by the substrate reduces the quasiparticle band gap by as much as 250~meV. 

\GWnot{} calculations yield an indirect fundamental band gap for the free-standing MoS$_2$ monolayer, using the experimental lattice parameters. Here, we found that in the presence of additional screening from the model  substrate, the \GWnot{} band gap exhibits a direct character. This result is independent of the frequency integration scheme (FF or PPA). The sensitivity of the direct/indirect character of the gap to substrate screening is an element to be taken into account when using \textit{ab initio} many-body calculations to predict the optoelectronic properties of 2D materials.

We also found that substrate screening affects the dispersion of quasiparticle bands. For example, screening enhances the electron and hole carrier effective masses at the {K} point by as much as 8\% and 27\%, respectively. The resulting masses are in very good agreement with experiments.

An analysis of the \GWnot{} self-energy and spectral function reveals that these results can be rationalized in terms of the shift of the plasma resonances as a result of the changing dielectric environment, in line with a simple Drude model of plasmon excitations.

On the methodology side, the calculations of interpolation-free quasiparticle effective masses and of spectral functions illustrate some of the capabilities of the SternheimerGW approach, and provide further validation of this emerging methodology. 

Our present findings provide insight into the role of the dielectric environment in the quasiparticle band structure of the prototypical TMD monolayer MoS$_2$. More generally, our work suggests that substrate engineering could offer new avenues to design future TMD-based electronic and optoelectronic devices.

\acknowledgments
We thank the MCC/Archer consortium (EP/L000202/1) and Isambard UK National Tier-2 HPC Service operated by GW4 and the UK Met Office, and funded by EPSRC (EP/P020224/1) for supercomputer resources.
We acknowledge PRACE for awarding us access to MareNostrum at the Barcelona Supercomputing Center (BSC), Spain.
F.G. was supported by the Computational Materials Sciences Program funded by the U.S. Department of Energy, Office of Science, Basic Energy Sciences, under Award DE-SC0020129.

\clearpage

\section{References}

\bibliographystyle{otago}
\bibliographystyle{apsrev}


\end{document}